# Hybrid adaptive splines for luminous intensity data regression in I-tables

L. Lipnický[1], R. Dubnička[1], J. Petržala[2] and L. Kómar[2,*]

[1] *Faculty of Electrical Engineering and Information Technology, Slovak University of Technology, Ilkovičova 3, Bratislava, Slovakia*

[2] *ICA, Slovak Academy of Sciences, Dúbravská cesta 9, Bratislava, Slovakia*

\* corresponding author: usarlako@savba.sk

## Abstract

The I-table contains luminous intensity values over the range of angles for the luminaires used in the road lighting in accordance with technical report CIE 121:1996. A limited number of angles causes smoothing of the luminous intensity diagram omitting possible local extremal values which affect the calculations of the photometric parameters such as average illuminance, average luminance, uniformity or treshold increment. The interpolating methods used to calculate the luminous intensity can significantly improve the accuracy of the calculations and redound to more effective and reliable road lighting design. In the paper standard interpolation methods used up to now are compared with newly proposed hybrid adaptive splines. Calculated values of luminous intensity are compared and verified by goniophotometric measurements.

## 1 Introduction

The purpose of the road lighting is to increase the safety and comfort of traffic participants at night. Road lighting provides good visibility conditions by illuminating the road and surrounding surfaces and by making objects visible to the drivers or pedestrians. Construction of road lighting can reduce night-time accidents by 20-40 %, especially when a participant of the car accident is a pedestrian. In the 1930s, the pioneer work of Waldram defined the silhouette principle of road lighting which states that targets on illuminated roads are seen as dark silhouettes against the bright road surface [1]. Since the late 1970s, systematic studies and improvements to the lighting of streets resulted in reduction of the crime in night-time.

The design of road lighting installation was shifted from that time towards the visible quantities, such as target luminance, road surface luminance, luminance uniformity, and restriction of glare [2]. A comprehensive analysis of 62 studies from 15 countries published by CIE (Commission Internationale de l'Éclairage - International Commision on Illumination) has the relevant answers for the illumination of the roads today. Several studies have tried to evaluate which photometric parameters (average illuminance, average luminance, uniformity of luminance, etc.) are important and how levels of these parameters are related to accident rate during night-time. Results of these studies have been collected and published as CIE document No.93, "Road lighting as an accident countermeasure" [3]. Nowadays, the road lighting design is based on CIE publications No. 115 "Lighting of roads for motor and pedestrian traffic" [4], and No. 140 "Road Lighting Calculations" published in 2019 [5].

Road lighting can improve the appearance of the environment at night, but care has to be taken to the upward light from the luminaires, which can form the sky glow, better known as light pollution [6]. In addition to making the astronomical observations of sky objects difficult or impossible, a lot of negative effects are reported on animals, vegetation and mankind [7]. Also an inappropriate lighting installation close to the housing can acts disturbingly to some residents. Therefore, correct quantification of the luminaire's photometric parameters play a decisive role in the road lighting design, and thus, the elimination of the unwanted light pollution effect.

The performance of an installation is assessed on basis of five characteristics, utilizing the most important photometric quantities: average luminance or illuminace of the road surface, the overall uniformity, the longitudinal uniformity, treshold increment and edge illuminance ratio. For acquiring and evaluation of these quantities, the luminous intensity distribution curve of the luminaires which are used in the road lighting must be known precisely. For this purpose, the so-called I-tables exist, which contain the luminous intensity values over the range of angles. In fact, the step in angles in the I-table is usually not sufficient for accurate calculations, especially for new LED luminaires with various optical elements and luminaires with faceted optical systems. Therefore, two different approaches are preferred; more precise goniophotometric measurements or fast and accurate interpolating methods. This paper proposes the usage of hybrid adaptive splines (HAS), known from other fields of science (e.g. geosciences [8], neurophysiology [9], economy [10], daylight engineering [11], etc.) to reconstruct the luminous intensity distribution curve from the I-table with arbitrary precision. The proposed method was compared with obviously used interpolating methods (linear, polynomial and cubic spline) in terms of I-table data reconstruction and verified by the goniophotometric measurements.

## 2 Theoretical background

The luminaires used in road lighting are characterized by the luminous intensities measured in *(C, γ)* coordinate system. They have specific shapes of the luminous intensity distribution curves because of their usage for lighting of a limited area in accordance with given possibilities of their location. The orientation of the C-planes is presented in Figure 1. As it can be seen, the plane C0-C180 is parallel with the road and the plane C90-C270 is perpendicular to the road. Every luminaire is therefore characterized by the luminous intensity distribution curve describing the luminous flux flowing through the space, and thus, the maximum luminous intensity, radiation angle, shading angle, etc. can be easily find out. Luminous intensity distribution curves are usually reported in the manufacturer's catalogs recalculated to 1000 lm. In the manufacturer's catalogs the luminous intensity distribution curves for road lighting are stated in the planes C0-C180 and C90-C270.

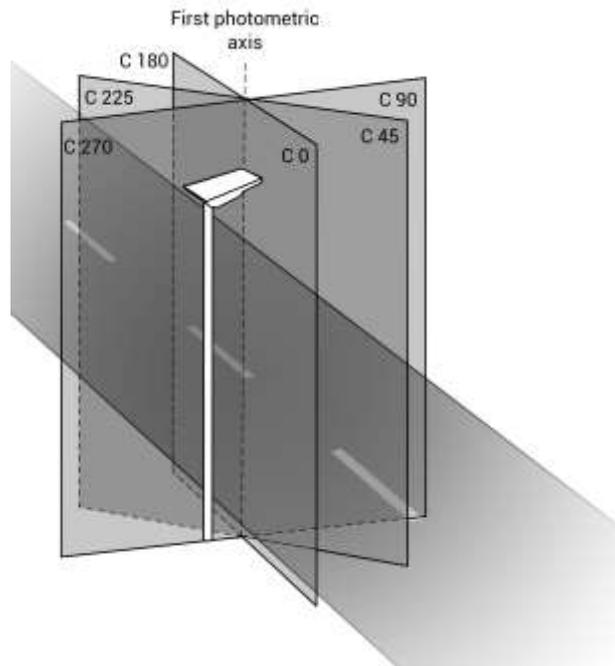

**Figure 1.** Orientation of the selected C-planes on a road.

Photometric parameters of the luminaire are inevitable for the quantitative characterization of the road lighting. The parameters are usually arranged in the I-tables in accordance with [12]. Nowadays, the file formats with a specific structure are recommended, such as EULUMDAT, IES, CIE, CEN, CIBSE TM14, etc. Given file formats are used in the appropriate software packages for road lighting design. Each producer offers a given electronic file format for its own luminaires, however, no unified file format exists which would eliminate deficiencies like omitting the temperature, light source location or the effect of the input voltage. The I-table contains the luminosity distribution of the luminaire to each relevant direction. In [5], the recommended angular resolution is maximally 5° for C-planes and 2.5° for $\gamma$. In fact, new luminaires with specific optical elements can have local extremal values of luminous intensities at the angles not listed in I-tables what can cause unwanted glare in result. Omitting this extremal values, the calculation of the photometric parameters is skewed and inaccurate. Therefore, the goniophotometric measurements are needed which can reach the angular resolution up to 0.5° – 1°. If it is necessary to estimate the values between the measuring points of the I-table, a fast and accurate interpolating method should be used.

## 3   Interpolation methods obviously used for luminous intensity calculation from I-table

Interpolation is a method of finding unknown function values using a given set of known values in surrounding points. If the new value has to be found from two given points then the linear interpolation formula is used, whereas if a set of numbers is available, polynomial fit or cubic spline can be used to find the value at a given coordinate. In general, the Lagrange polynomial has the

form:

$$y(x) = \sum_{i=1}^{n} y_i \cdot \prod_{j=1, j \neq i}^{n} \left( \frac{x - x_j}{x_i - x_j} \right), \quad (1)$$

where *y(x)* is the interpolated value at a point *x* and *n* is the total number of points $x_i$ with known values $y_i$ between which we interpolate. The most frequently used interpolating method in road lighting calculations is the linear interpolation. Also the quadratic interpolation was allowed in older publication [13]. Selection of the interpolating method strongly depends on the accuracy as well as on the necessary CPU time. Note, that there is a 2D array in all I-tables, thus an interpolation in both dimensions is required which can significantly extend a time for the mass calculations.

### 3.1 Linear interpolation

To obtain a value of the luminous intensity *I(C, γ)* lying between the measured values in the I-table four neighboring measured values of *I* are needed. The well-known equation for linear interpolation was used in the form:

$$y(x) = y_1 \left( \frac{x - x_2}{x_1 - x_2} \right) + y_2 \left( \frac{x - x_1}{x_2 - x_1} \right). \quad (2)$$

The coordinates *x, $x_1$, $x_2$* are at first replaced by the azimuth angle *C* and then by the photometric angle *γ*. The order of *C* and *γ* will not affect the results. If coordinates *x, $x_1$, $x_2$* are at first replaced by the azimuthal angle *C* then the terms *I(C, $γ_j$)* and *I(C, $γ_{j+1}$)* are obtained. Consequently, using this terms we obtain a searched value of the luminous intensity *I(C, γ)* in accordance with [5]:

$$I(C, \gamma) = I(C, \gamma_j) + \frac{\gamma - \gamma_j}{\gamma_{j+1} - \gamma_j} \times \left[ I(C, \gamma_{j+1}) - I(C, \gamma_j) \right]. \quad (3)$$

### 3.2 Polynomial interpolation of higher degree

A commonly used procedure in various fields of science is to find the unknown value between data point in form of the polynomial of the higher degree. Note that the polynomial interpolation of the first degree corresponds to linear interpolation. Polynomial interpolation relies on the fact that there is a unique polynomial of degree *n* that passes through these *n+1* data points. To find the unique polynomial, we need to solve a set of *n+1* equations with the unknown coefficients. Let *($X_i$, $y_i$)* be the $i^{th}$ data point where *i < (n+1)*, $X_i$ is a vector given by a set of coordinates in I-table as follows: $X_i$ = *($C_m$, $γ_j$)*, $y_i$ = *I($C_m$, $γ_j$)*. Then, we have a set of equations:

$$a_0 + a_1 X_1 + a_1 X_1^2 + \ldots + a_n X_1^n = y_1$$
$$a_0 + a_1 X_2 + a_1 X_2^2 + \ldots + a_n X_2^n = y_2$$

$$a_0 + a_1 X_{n+1} + a_1 X_{n+1}^2 + \ldots + a_n X_{n+1}^n = y_{n+1} \quad (4)$$

which can be solve in a matrix form using Gaussian elimination method to obtain the unknown coefficients $a_0 \ldots a_n$. There are various disadvantages of the interpolating polynomials method. The major disadvantage is that as the number of data points increases, the order of the interpolating polynomial increases as well. However, for large degree ($n > 10$), the polynomial would be oscillating wildly between the data points leading to large errors in the interpolated values. As it will be shown in this paper, this method scarcely can be used for data reconstruction in I-tables.

### 3.3 Cubic spline

Linear interpolation has some loss in accuracy, if larger angular intervals are used and polynomial interpolation of higher degree has a well-known inaccuracy for huge data set. Therefore, cubic spline is derived as follows:

$$S_i(x) = a_i + b_i(x - x_i) + c_i(x - x_i)^2 + d_i(x - x_i)^3. \quad (5)$$

We are looking for the coefficients of following polynomial system to find the value of $C$ at an angle of $\gamma_{j-1}$.

$$S_0(C) = a_0 + b_0(C - C_{m-1}) + c_0(C - C_{m-1})^2 + d_0(C - C_{m-1})^3,$$

$$S_1(C) = a_1 + b_1(C - C_m) + c_1(C - C_m)^2 + d_1(C - C_m)^3,$$

$$S_2(C) = a_2 + b_2(C - C_{m+1}) + c_2(C - C_{m+1})^2 + d_2(C - C_{m+1})^3. \quad (6)$$

Conditions for a natural cubic spline can be written as:

$$S_0(C_{m-1}) = 0 \Rightarrow 2c_0 = 0,$$

$$S_2(C_{m+2}) = 0 \Rightarrow 2c_2 + 6d_2(C_{m+2} - C_{m+1}) = 0. \quad (7)$$

The coefficients $b_i, c_i, d_i$ are calculated using the following formulas:

$$h_i c_i + 2(h_{i+1} + h_i) c_{i+1} + h_{i+1} c_{i+2} = 3\left(\frac{\Delta y_{i+1}}{h_{i+1}} - \frac{\Delta y_i}{h_i}\right),$$

$$b_i = \frac{\Delta y_i}{h_i} - \frac{h_i}{3}(c_{i+1} + 2c_i),$$

$$d_i = \frac{c_{i+1} - c_i}{3h_i}. \tag{8}$$

Therefore, it is needed to repeat this procedure for angles $\gamma_j$, $\gamma_{j+1}$, $\gamma_{j+2}$ to find the interpolated value $I(C, \gamma)$.

## 4  Proposal of hybrid adaptive splines for I-tables

### 4.1  Smoothing splines

When observational data are loaded by measurement errors, we propose an unknown function which does not coincide with given data points exactly. To obtain the unknown function based on noisy data we use the method of smoothing splines first. It is a trade-off between fidelity to the data and a requirement of adequate smoothness of a so-called regression function. In general, we can suppose a function $f$ of two variables $C_m$ and $\gamma_j$ which correspond to the coordinate system defined in I-table:

$$w_k(C_m, \gamma_j) = f_k(C_m, \gamma_j) + \varepsilon_k \tag{9}$$

where $w_k(C_m, \gamma_j)$ $k = 1, ..., N$ are the measured values of the luminous intensity corresponding to the $N$ data points $(C_m, \gamma_j)$ and $\varepsilon_k$ is zero-mean independent random error with a common variance $\sigma^2$. We are looking for a such regression function $f$ which minimizes the spline functional

$$F_{p,\lambda}(f) = \frac{1}{N}\sum_{k=1}^{N}\left(w_k(C_m, \gamma_j) - f_k(C_m, \gamma_j)\right)^2 + \lambda J_p(f), \tag{10}$$

where $J_p$ is in general some functional which secures smoothness of $f$ up to an "order" $p$ ($p$ is an integer $\geq 2$). The smoothing parameter $\lambda$ controls the trade-off between smoothness of the regression function and fidelity to the data. The optimal values of the parameters $\lambda$ and $p$ is determined from measured data.

In our case, we try to estimate a luminous intensity on the sphere, thus we will use the method of smoothing spline on the sphere as can be found in [14]. We have a set of measured values $w_k(C_m, \gamma_j)$ in points $(C_m, \gamma_j)$, where $\gamma \in \langle 0, \pi \rangle$ is a spherical latitude and $C \in \langle 0, 2\pi \rangle$ is a spherical longitude. The functional $J_p$ now has the form

$$J_p(f) = \int_0^{2\pi}\int_0^{\pi}\left[\Delta^{p/2} f(C, \gamma)\right]^2 \sin\gamma \, d\gamma \, dC \tag{11}$$

for $p$ even, and

$$J_p = \int_0^{2\pi} \int_0^{\pi} \left\{ \frac{1}{\sin^2\gamma} \left[ \frac{\partial}{\partial C} \left( \Delta^{\frac{p-1}{2}} f(C,\gamma) \right) \right]^2 + \left[ \frac{\partial}{\partial \gamma} \left( \Delta^{\frac{m-1}{2}} f(C,\gamma) \right) \right]^2 \right\} \sin\gamma \, d\gamma \, dC \quad (12)$$

for *p* odd. The term $\Delta$ represents the Laplace operator, which on the sphere has the form

$$\Delta \equiv \frac{1}{\sin^2\gamma} \frac{\partial^2}{\partial C^2} + \frac{1}{\sin\gamma} \frac{\partial}{\partial \gamma} \left( \sin\gamma \frac{\partial}{\partial \gamma} \right). \quad (13)$$

Now, let us use a standard unit direction vector defined as $\vec{e} = (\cos C \sin\gamma, -\sin C \sin\gamma, \cos\gamma)$ with its beginning in the center of the sphere for determining of a point's position *(C, γ)* on the sphere. We can write the solution of the minimization problem of the functional (11)-(13) in the form

$$f_{p,\lambda}(\vec{e}) = \sum_{k=1}^{N} c_k K(\vec{e}, \vec{e}_k) + d,$$

$$K(\vec{e}, \vec{e}_k) = \frac{1}{4\pi} \sum_{\nu=1}^{\infty} \frac{2\nu+1}{[\nu(\nu+1)]^p} P_\nu(\vec{e} \cdot \vec{e}_k) \quad (14)$$

where $P_\nu$ are the Legendre polynomials. For more detailed derivation, see [11]. The function $f_{p,\lambda}$ represents the solution of the minimalization problem for some given values of *p* and $\lambda$, and thus, the goal to find a searched value of the luminous intensity in the I-table. The criterion for a good choice of the smoothing parameter $\lambda$ and "the order of smoothness" *p* is the ability to predict the value of the investigated field beside the known data points. The typical method for the parameters optimization is the method of generalized cross-validation (see e.g. [15] or [16]).

### 4.2 Hybrid adaptive splines

When sharp local peaks appear in measured data, some adaptive spline method must be used to estimate the true function. It can be done by some smoothing spline method with locally variable smoothing parameters or by a regression spline method with adaptive placing of knots (or, equivalently, adaptive selection of spline basis functions). We will use the method of hybrid adaptive splines (HAS) proposed by [17] which combines features of adaptive regression splines and smoothing splines on sphere presented above.

In the case of smoothing splines on the sphere, we was looking for a regression function minimizing the functional $F_{p,\lambda}(f)$ generated by the basis functions $K(\vec{e}, \vec{e}_i)$, $i = 1, ..., N$. The HAS method is based on selection of a convenient subset of these basis functions and minimizing of the functional $F_{p,\lambda}(f)$ in the function space generated by this subset. When we were chosen the optimal system of *(M+1)* basis functions (method is described in [11]), the regresion function will have a form

$$f_{p,\lambda}(\vec{e}) = d + \sum_{l=1}^{M} c_l K(\vec{e}, \vec{e}_{i_l}). \quad (15)$$

For the coefficients $d$ and $(c_1,...,c_M)^T = \vec{c}$ the following expressions hold

$$d = \left(LH^{-1}K_1^T\vec{T}\right)^{-1} LH^{-1}K_1^T\vec{w}_k,$$
$$\vec{c} = H^{-1}K_1^T\left(\vec{w}_k - \vec{T}d\right),\tag{16}$$

where

$$H = K_1^T K_1 + N\lambda K_2,$$
$$L = \vec{T}^T \left(K_1 K_1^T\right)^{-1} K_1 K_2$$
$$K_1 = K\left(\vec{e}_k, \vec{e}_{i_l}\right), k=1,...,N, l=1,...,M$$
$$K_2 = K\left(\vec{e}_{i_k}, \vec{e}_{i_l}\right), k=1,...,M, l=1,...,M$$
$$\vec{T} = \begin{pmatrix} 1 \\ \vdots \\ 1 \end{pmatrix}_{N\times 1},\tag{17}$$

As in the case of the smoothing splines we must optimize the free parameters $p$ and $\lambda$. The convenient values of the parameters are determined by minimizing of the generalized cross-validation function analogically as in the case of the smoothing splines.

The procedure of HAS seems to be complicated and non-effective, but the routine calculations on a standard PC prove its speed and high accuracy, especially if significant non-homogeneities occure in measured data in the I-tables. MATLAB code for HAS calculation as well as calculation of rRMSE and $R^2$ for interpolating methods discussed in this paper is freely available on webpage: www.skyglow.sav.sk.

## 5 Numerical examples and validation

Three luminaires with different luminous intensity distribution curves obtained from the manufacturer have been selected for the purpose of the accuracy testing of the described interpolation methods. The relative root mean square error (rRMSE) and coefficient of determination ($R^2$) were chosen as appropriate statistical indicators to compare the calculated data with the goniophotometric measurement with the step 2.5 degree in the C-planes and 0.5 degree in $\gamma$-angles. For the road lighting purposes data in the I-tables are commonly tabled with the step of 5 degree in the C-planes and 2.5 degree in $\gamma$-angles. Therefore, we compared the accuracy of the interpolation methods to the experimentally obtained data for this frequently used grid.

|  | Luminaire 1 | | Luminaire 2 | | Luminaire 3 | |
| --- | --- | --- | --- | --- | --- | --- |
|  | rRMSE (%) | $R^2$ | rRMSE (%) | $R^2$ | rRMSE (%) | $R^2$ |
| **Linear** | 7.46 | 0.9967 | 6.22 | 0.9982 | 14.16 | 0.9846 |
| **Polynomial** | 43.75 | 0.8740 | 49.54 | 0.8733 | 43.99 | 0.8290 |
| **Cubic spline** | 6.99 | 0.9971 | 6.07 | 0.9983 | 14.08 | 0.9848 |
| **HAS** | 2.21 | 0.9995 | 2.61 | 0.9993 | 4.23 | 0.9982 |

**Table 1.** The relative root mean square error (rRMSE) and coefficient of determination ($R^2$) for various interpolating methods. Input data are for the grid with step in C-plane 5º and in γ-plane 2.5º and calculation grid corresponds to step in C-plane 2.5º and in γ-plane 0.5º.

As can be seen from Table 1, the polynomial fit of higher degree ($n = 10$) is far beyond tolerable accuracy, and therefore it is not recommended to use it to reconstruct the missing I-table data. The linear interpolation is realistically accurate if we meet the rRMSE range of 6-14% depending on the luminaire. Luminaire 3 has the most sharp peaks in luminous intensity distribution, so this luminaire has a higher error than other luminaires when reconstructing data with linear interpolation. However, this type of interpolation responds very slowly to sudden fluctuations in luminous intensity value and it is therefore more suitable for reconstructing more shallow luminous intensity distributions (like in cases of Luminaire 1 and 2). Also, the examined cubic spline shows a slightly lower rRMSE in Luminaire 3 than linear interpolation, therefore it seems to be more accurate when reconstructing luminous intensity distribution without the presence of significant local extremes. HAS is far more accurate on average compared to other interpolation methods, with rRMSE not exceeding 4.3% even with Luminaire 3 with more local peaks. This is due to the adaptive selection of local base functions in nodes that are sufficiently surrounded by known data.

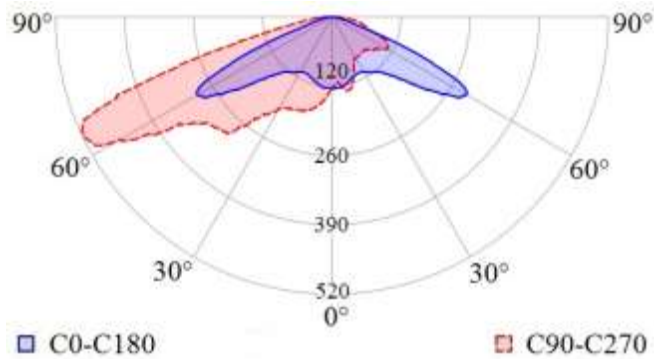

**Figure 2.** Goniophotometric mesaurement of luminous intensity distribution curve of the Luminaire 1 in planes (C0-C180 and C90-C270).

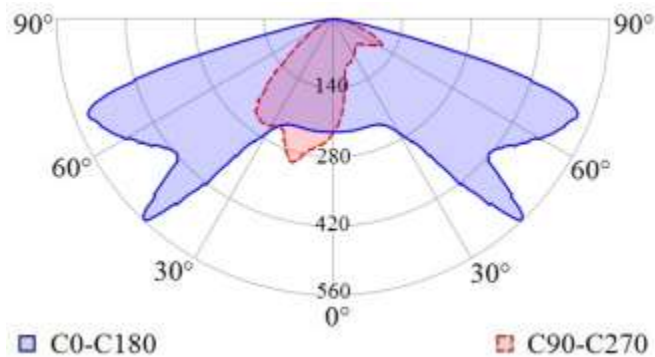

**Figure 3.** Goniophotometric mesaurement of luminous intensity distribution curve of the Luminaire 2 in planes (C0-C180 and C90-C270).

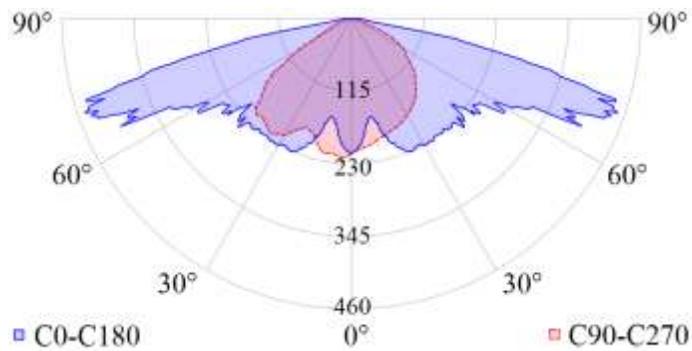

**Figure 4.** Goniophotometric mesaurement of luminous intensity distribution curve of the Luminaire 3 in planes (C0-C180 and C90-C270).

Figures 2 - 4 shows the luminous intensity distribution curves of the tested luminaires in the C0-C180 and C90-C270 planes which were measured in the laboratory. These curves were compared with the curves reconstructed by selected interpolation methods from input values measured with steps in C-plane = 2.5° and $\gamma$-plane = 0.5°.

Figure 5 shows the luminous intensity distribution of the luminaire in the polar coordinates given by the planes C in the range of 0 - 360 degrees and the plane $\gamma$ in the range of 0 - 80 degrees due to the most significant changes at just 80 degrees. In Figure 5b, the polynomial fit of $10^{th}$ order clearly distorts the luminous intensity values while the linear interpolation and HAS give more or less realistic values, whereas the HAS (Figure 5c) better describes the distinct peaks in luminous intensity values.

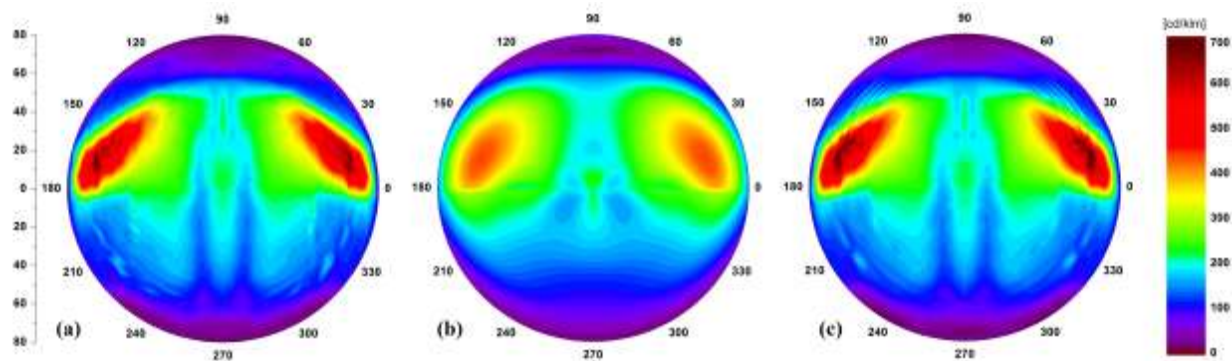

**Figure 5.** Luminance distribution for Luminaire 3 calculated by linear interpolation (a), polynomial of $10^{th}$ degree (b) and HAS (c).

Figure 6 shows the relative deviations between the calculated luminous intensity values using four interpolation methods and data measured by a goniophotometer with steps in C-plane = 2.5° and $\gamma$-plane = 0.5° for Luminaire 3. Obviously, the polynomial fit (Figure 6b) gives deviations > 20%, especially in the parts of increased luminance as well as at the edges. It is therefore extremely unsuitable for reconstructing data in the I-table.

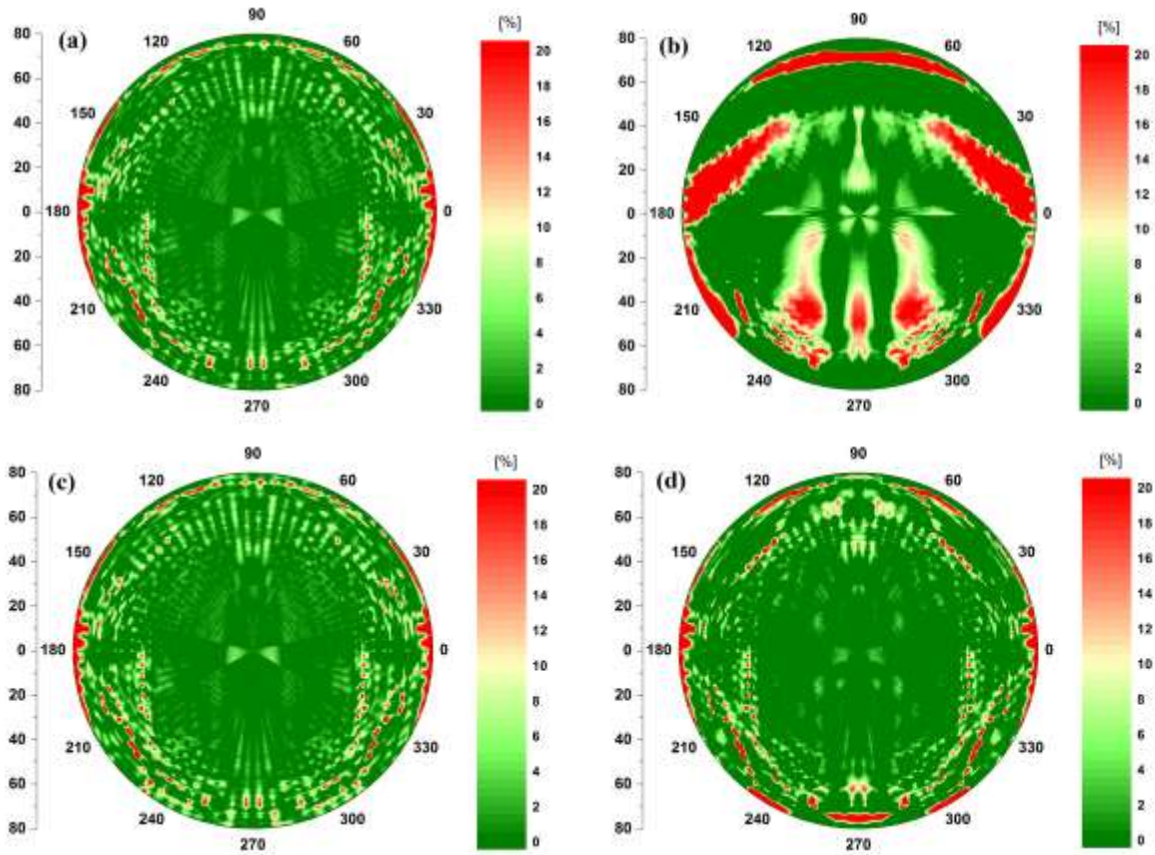

**Figure 6.** Realtive deviation between measured and calculated luminances for Luminaire 3 and four interpolating methods: (a) – linear, (b) – polynomial of 10$^{th}$ degree, (c) – cubic spline and (d) – HAS on 2.5º x 0.5ºgrid.

The linear interpolation (Figure 6a) and cubic spline (Figure 6c) show very similar relative deviation distributions, but on average the cubic spline has lower deviations. This also corresponds to a slightly lower rRMSE for cubic spline (see Table 1). In the case of HAS (Figure 6d), the relative deviation is the lowest in the middle of the graph, i.e. for angles in γ-plane up to 45 degrees, with peaks being reconstructed within the deviation of 10%. The increase in relative margin in Fig. 6d ($γ > 65º$) is due to the less data needed to create the interpolation model. However, the HAS method reliably reconstructs I-table data for smaller angles in γ-plane even in a sudden change in luminous intensity values. For angles $γ > 65º$, it is recommended to use linear interpolation or cubic spline in case of a denser grid in road lighting calculation, respectively.

## 6  Conclusions

The presented paper deals with various interpolation methods used in the reconstruction of data from I-table, their comparison with goniophotometric measurements, as well as testing of hybrid adaptive splines that have not been used in this field so far. Linear interpolation, 10$^{th}$ degree polynomial fit, cubic spline and hybrid adaptive spline were compared. As the results show, the so far used linear interpolation provides sufficient accuracy in data reconstruction, but relatively slowly responds to sudden changes in luminous intensity distribution. This reduces its accuracy, especially for luminaires with local extremes, which can cause glare of road users. Cubic spline shows little higher accuracy than linear interpolation, but fails again at local extremes. Therefore, the HAS method was designed to adaptively respond to local extremes in luminous intensity distribution and it provides the highest match compared to the measured data. However, this method fails in the edges where the number of data is limited to create an accurate interpolation model. It is therefore questionable under what conditions to use this method to ensure sufficient accuracy and to identify possible angles that may cause glare. Ultimately, it can be stated that the cubic spline method is recommended to be used in all cases of flat luminous intensity distribution, also limited to curves with local extremes, where it exhibits comparable accuracy as linear interpolation. In the case of a redundant network of points it is recommended to use HAS, especially at angles < *65* degrees in $\gamma$-plane. The polynomial fit was tested only to make sure it was an inappropriate method and should not be used to reconstruct the data from the light curve at all.

## Acknowledgements

This work was supported by the Slovak Research and Development Agency under the contract No. APVV-18-0014 and by the Grant Agency VEGA under the contract number No. 1/0640/17.